\documentclass[twocolumn]{article}
\usepackage[T1]{fontenc}
\usepackage{helvet}
\usepackage{mathptmx}
\usepackage[labelfont=bf,labelsep=period]{caption}
\usepackage[margin=0.75in, columnsep=0.25in]{geometry}
\usepackage{siunitx}
\usepackage{amsmath,amsfonts,amsthm,amssymb}
\usepackage[blocks]{authblk}

\usepackage{graphicx}
\usepackage{xcolor}
\usepackage[style=chem-acs, backend=biber, doi=true, autocite=superscript]{biblatex}
\definecolor{acsblue}{RGB}{13, 84, 165}
\definecolor{acsgreen}{RGB}{0, 69, 56}
\usepackage[colorlinks=true,linkcolor=acsblue,citecolor=acsblue,urlcolor=acsblue]{hyperref}
\usepackage{titlesec}
\titleformat{\section}{\normalfont\sffamily\large\bfseries\color{acsgreen}}{\thesection}{0.5em}{}
\titleformat{\subsection}{\normalfont\sffamily\bfseries\color{acsgreen}}{\thesubsection}{0.5em}{}
\usepackage{orcidlink}
\usepackage{microtype}
\begin{document}

\author{Jin Zhang$^*$\orcidlink{0000-0003-4982-1830} and Peter W. Voorhees\orcidlink{0000-0003-2769-392X}}
\affil{Department of Materials Science and Engineering, Northwestern University, Evanston, IL 60208, United States}

\title{\sffamily\bfseries Morphological Stability of Metal Anodes: Roles of Solid Electrolyte Interphases (SEIs) and Desolvation Kinetics}
\date{}

\twocolumn[
\maketitle
\vspace{-2em}
\noindent
\mbox{\begin{minipage}[t]{0.58\textwidth}
\vspace{0pt}
\small{\color{acsgreen}\textbf{ABSTRACT:}} Achieving stable lithium metal anodes requires control over the solid-electrolyte interphase (SEI) and desolvation kinetics. Here, we develop a unified theoretical framework integrating ion transport, desolvation, charge transfer, and SEI breakdown to predict morphological instabilities during electrodeposition. Using linear stability analysis, we identify six dimensionless parameters that govern the onset and evolution of instabilities. We show that SEI transport and desolvation rate effectively modulate apparent reaction kinetics, shifting the system toward a stable, reaction-limited regime. Extending the classical limiting current concept, we demonstrate that a thick, poorly conductive SEI and sluggish desolvation significantly reduce the limiting current. We introduce an apparent Damk\"ohler number to quantify the critical balance: suppressing diffusion-limited instabilities by reaction rate reduction, while maintaining a high limiting current. Our theory enables predictive mapping of electrodeposition morphologies across diverse materials and operating conditions, guiding the rational design of stable lithium metal anodes.
\end{minipage}}%
\hfill
\mbox{\begin{minipage}[t]{0.38\textwidth}
\vspace{0pt}
\centering
\includegraphics[width=\textwidth]{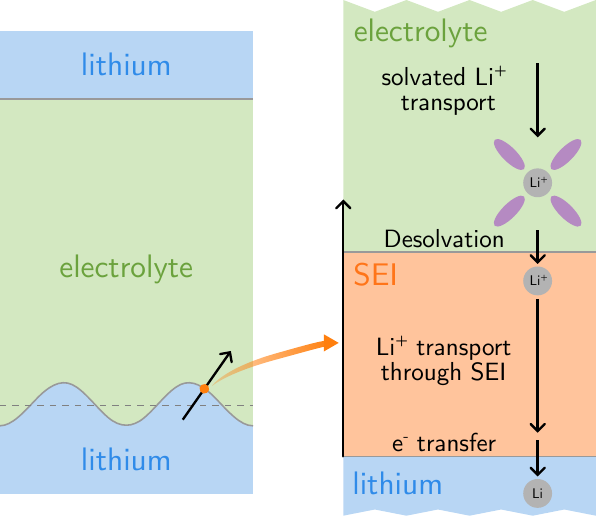}
\end{minipage}}
\vspace{2em}
]
\let\thefootnote\relax\footnotetext{$^*$Email: \href{mailto:jzhang@northwestern.edu}{jzhang@northwestern.edu}, \href{mailto:zhang.jin@outlook.com}{zhang.jin@outlook.com}\\This document is the Accepted Manuscript version of a Published Article that appeared in final form in ACS Energy Letters, copyright \copyright~2026 The Authors, Published by American Chemical Society. To access the final published article, see \url{https://pubs.acs.org/doi/10.1021/acsenergylett.5c03690}.}

Lithium metal is a promising anode material for next-generation batteries. However, its practical use is hindered by the intrinsic instability arising from lithium's high reactivity with liquid electrolytes, leading to the formation of a solid electrolyte interphase (SEI) at the electrode interface. This SEI layer plays a critical role in battery performance \autocite{Hobold2021}. During electrodeposition, solvated lithium ions must desolvate, transport through the SEI, and finally deposit onto the lithium surface. These processes are governed by a complex interplay among electrolyte diffusion and migration, SEI transport, desolvation, and charge transfer, which can be optimized through electrolyte engineering and artificial SEI design \autocite{Xu2004,Lopez2019,Huang2025}. Despite their importance to the efficiency, reversibility, and safety of lithium metal anodes, the coupled mechanisms governing interfacial stability remain poorly understood.

Recent advances in experimental techniques have revealed unprecedented detail on SEI's structure and composition. Cryoelectron microscopy (cryo-EM) enables direct imaging of SEI nanostructures and their early-stage formation \autocite{Li2017,Li2018,Wang2025}. Complementing these structural insights, carefully designed experiments are used to quantify SEI properties such as Li$^+$ diffusivity, active Li$^+$ concentration \autocite{Boyle2022}, and transference number \autocite{Liu2025}. Solvation structure has long been used in electrolyte engineering to tailor bulk transport, desolvation kinetics, and SEI properties \autocite{Kim2021,Kim2025,PlazaRivera2025}. A long-standing goal of battery research has been to correlate SEI and solvation properties with battery performance, particularly Coulombic efficiency (CE). Several metrics have been proposed for this purpose. Hobold et al. \autocite{Hobold2023} introduced the pseudo/apparent exchange current density ($j_0^p$) to quantify the effective reaction of SEI-covered Li, observing a positive correlation between $j_0^p$ and CE. Metrics balancing kinetics and diffusion, e.g., $j_0^p/FcD$ \autocite{Hobold2021} and $j_0^p/D$ \autocite{PlazaRivera2025} ($D$ and $c$ are electrolyte diffusivity and concentration, respectively) have also generally shown a positive correlation with CE. In terms of the second Damk\"ohler number (Da) (the ratio of reaction to diffusion rates), these correlations seemingly suggest that a higher Da is beneficial. This contrasts with theoretical \autocite{Cogswell2015,Enrique2017,Khoo2019,Zheng2021,Zhang2025} and experimental \autocite{Chen2020,Choudhury2023,Jin2024} findings that a smaller Da is preferred to achieve a reaction-limited regime that suppresses diffusion-related instabilities. This contradiction remains unresolved. Additional correlations include CE with SEI cT number \autocite{Liu2025}, and improved CE with less negative Li$^+$/Li potential or weaker solvation \autocite{PlazaRivera2025}. Together, these findings highlight the urgent need for a unified theoretical framework that systematically connects SEI and electrolyte transport, charge-transfer, desolvation kinetics, and interfacial stability under realistic operating conditions.

\begin{figure*}[!t]
  \centering
  \includegraphics[width=1.0\linewidth]{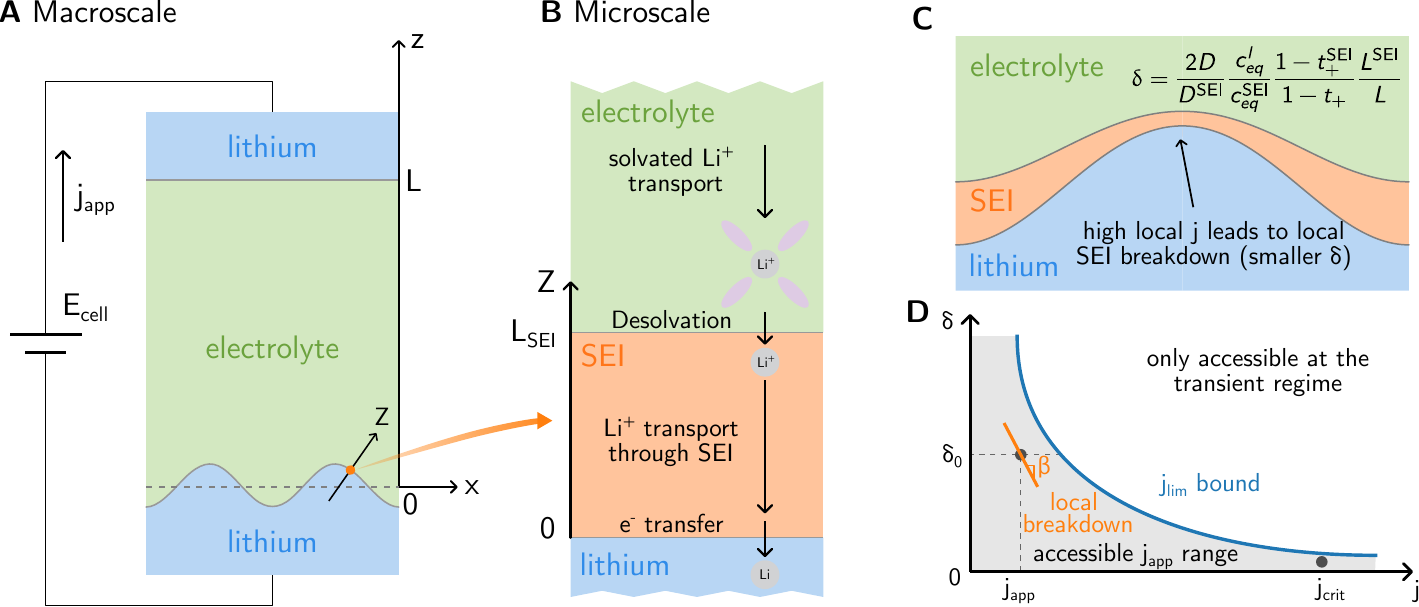}%
  \caption{Multiscale modeling of lithium metal anodes. (A) Schematic of the macroscale symmetric cell. (B) Microscale depiction of ion transport and kinetics, illustrating sequential steps of bulk transport, desolvation, SEI transport, and charge transfer at the lithium interface. (C) Physical picture of local SEI breakdown: high local current density leads to a reduction in SEI thickness ($L_{\text{SEI}}$), increased effective diffusivity $D_{\text{SEI}}$ and active lithium concentration $c_{\text{SEI}}$, which collectively reduce the SEI parameter $\delta$. (D) SEI breakdown rate and operational constraints. Applied current density ($j_{\text{app}}$) is bound by the limiting current density ($j_{\text{lim}}$). The SEI breakdown rate is quantified by the SEI breakdown parameter $\beta$.}\label{fig:schematic}
\end{figure*}

Here, we present a comprehensive theoretical framework integrating four key physical processes: (1) a multiscale approach bridging nanoscale SEI and macroscale electrolyte behavior; (2) desolvation at the SEI-electrolyte interface and charge transfer at the electrode-SEI interface; (3) distinct ionic transport in the electrolyte and SEI; and (4) SEI breakdown during cycling. We consider a symmetric cell configuration as shown in Fig.~\ref{fig:schematic}. The nanoscale SEI thickness results in a clear length scale separation from the separator. The multiscale ion transport is captured by coupling an SEI model (Fig.~\ref{fig:schematic}B) to every macroscopic electrode-electrolyte interfacial point (Fig.~\ref{fig:schematic}A), linking the respective electrode-SEI and SEI-electrolyte interfaces. In the bulk electrolyte, solvated Li$^+$ ions transport via diffusion and migration, as shown in Fig.~\ref{fig:schematic}B. Upon reaching the electrolyte-SEI interface, Li$^+$ undergoes desolvation before transporting across the SEI to the lithium metal anode, where it is reduced and incorporated into the metal lattice. The overall battery performance depends critically on the efficiency of each step.

We consider a representative electrolyte: 1 M LiPF$_6$ in EC:DEC 3:7 v/v, with material parameters summarized in Table~S1. Our framework is generalizable to other electrolyte-metal systems. We denote the ion concentration and electric potential in the electrode, SEI, and electrolyte phases as $c^s$, $c^{\text{SEI}}$, $c^l$, and $\phi^s$, $\phi^{\text{SEI}}$, $\phi^l$, respectively. The negative electrode interface is assumed to be perturbed and evolves as $h(x,t)$, while the positive electrode (at $z=L$) remains planar, enabling focus on single-interface stability. Asymptotic analysis (Supporting Information) allows treating the SEI problem as planar in a local coordinate $Z$ aligned with the macroscale interface normal (Fig.~\ref{fig:schematic}A). Ion transport in both the electrolyte and SEI is governed by diffusion and migration, described by the Nernst-Planck equation \autocite{Newman2021}. Charge transfer at the electrode-SEI interface is described by the symmetric Butler-Volmer kinetics: $j = -2 j_0 \sinh \frac{F}{RT} \frac{\eta}{2}$, with $j_0$ the intrinsic exchange current density and $\eta$ the overpotential. Desolvation at the SEI-electrolyte interface is modeled by a linearized Butler-Volmer kinetics: $j = -j_{0,\text{solv}} \frac{F}{RT}\eta_{\text{solv}}$, with $j_{0,\text{solv}}$ the exchange current density for solvation/desolvation, and $\eta_{\text{solv}}$ the solvation overpotential.

The limiting current ($j_{\text{lim}}$), which governs the maximum sustainable current density under quasi-steady state, is critical to battery performance. It is defined by complete lithium-ion depletion. In the presence of the SEI, the limiting condition should be defined by the SEI lithium concentration at the electrode-SEI interface, not the traditional electrolyte value at the SEI-electrolyte interface. The Li$^+$ concentration at the electrode interface ($Z=0$), incorporating SEI and desolvation, is (derivation in Supporting Information):
\begin{equation}
  \label{eq:cSEI0}
  c^{\text{SEI}}(Z=0) = A \left( 1 - \frac{j}{j_{\text{lim}}^c} \left(1+\delta \exp \left(-\frac{F}{RT}\frac{\eta_{\text{solv}}(j)}{2}\right)\right)\right),
\end{equation}
where $A$ is defined in Eq.~S50, $j_{\text{lim}}^c= \frac{2 F D c_0}{(1-t_+)L}$ denotes the classical limiting current in the absence of SEI. Similarly, in the SEI-dominated limit, we define SEI limiting current as $j_{\text{lim}}^{\text{SEI}}=\frac{F D^{\text{SEI}} c_{\text{eq}}^{\text{SEI}}}{(1-t_+^{\text{SEI}})L^{\text{SEI}}}\frac{c_0}{c_{\text{eq}}^l}$. Symbol definitions are in Table~S1. To quantify SEI effects, we introduce a dimensionless SEI parameter:
\begin{equation}
  \label{eq:delta}
  \delta = \frac{2D}{D^{\text{SEI}}}\frac{c_{\text{eq}}^l}{c_{\text{eq}}^{\text{SEI}}}\frac{1-t_+^{\text{SEI}}}{1-t_+}\frac{L^{\text{SEI}}}{L}.
\end{equation}

\begin{figure*}[!t]
  \centering
  \includegraphics[width=0.88\linewidth]{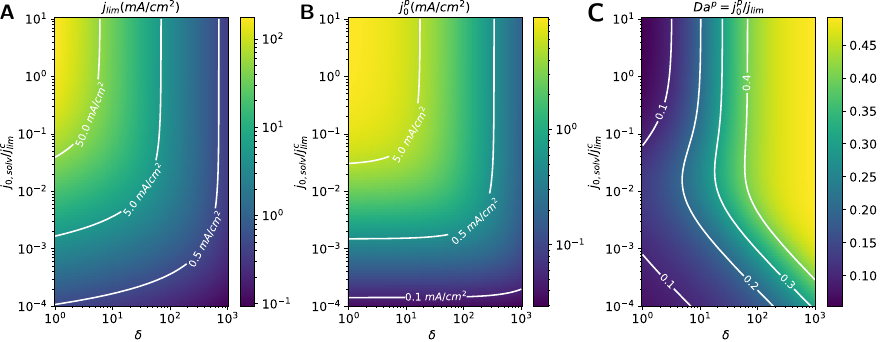}%
  \caption{Effect of the SEI parameter $\delta$ and the desolvation rate (quantified by the dimensionless solvation exchange current density $j_{0,\text{solv}}/j_{\text{lim}}^c$) on (A) limiting current density $j_{\text{lim}}$; (B) apparent exchange current density $j_0^p$; and (C) apparent Damk\"ohler number $\text{Da}^p$.}\label{fig:jlim}
\end{figure*}

Here, $\delta$ represents the ratio of the diffusion-limited currents for the electrolyte-dominated and SEI-dominated regimes: $\delta=j_{\text{lim}}^c/j_{\text{lim}}^{\text{SEI}}$. The actual limiting current ($j_{\text{lim}}$) lies between these two limits. Smaller $\delta$ and faster desolvation kinetics increase $j_{\text{lim}}$ (Fig.~\ref{fig:jlim}A). In the limit of infinitely fast desolvation ($\eta_{\text{solv}}=0$), the limiting current expression reduces to $j_{\text{lim}} = j_{\text{lim}}^c/(1+\delta)$, indicating an inverse dependence on $\delta$. From Table~S1, we calculate $\delta=386$ and $j_{\text{lim}}=0.91~\si{\mA/\cm^2}$. This indicates an SEI-dominated regime ($\delta\gg 1$) where ionic transport through the SEI layer becomes rate-limiting, significantly reducing the limiting current from the electrolyte-dominated value of $j_{\text{lim}}^c=353~\si{\mA/\cm^2}$. This is consistent with ultramicroelectrode measurements \autocite{Boyle2020}, which show that SEI transport is far more resistive than intrinsic charge transfer.

The role of SEI on the effective reaction rate can be captured by an apparent exchange current density $j_0^p$ \autocite{Hobold2023}, obtained from the cell voltage $E_{\text{cell}}$ for a symmetric cell:
\begin{equation}
  \label{eq:j0p}
  \frac{1}{j_0^p} = \frac{1}{2}\frac{F}{RT} \left.\frac{\mathrm{d} E_{\text{cell}}}{\mathrm{d} j}\right|_{j=0} = \frac{1}{j_{0}} + \frac{1}{j_{0,\text{solv}}} + \frac{2(1+\delta)}{j_{\text{lim}}^c}.
\end{equation}
In Eq.~\ref{eq:j0p}, the three terms on the right-hand side account for intrinsic charge transfer, desolvation, and ion transport across the electrolyte and SEI. Results in Fig.~\ref{fig:jlim}B are consistent with experimental findings that weaker solvation (faster desolvation, $j_{0,\text{solv}}$) increases $j_0^p$ \autocite{Boyle2022a}, while the presence of SEI ($\delta>0$) leads to lower exchange current density \autocite{Boyle2020}. The measured exchange current density from ultramicroelectrode techniques is likely a convolution of charge transfer and desolvation: $j_0^{\text{UME}} = (1/j_0+1/j_{0,\text{solv}})^{-1}$. Using parameters from Table~S1 and the $\delta$ value from Eq.~\ref{eq:delta}, we compute from Eq.~\ref{eq:j0p} $j_0^p=0.44~\si{\mA/\cm^2}$, which is in excellent agreement with the experimental range of $0.26$ - $0.55~\si{\mA/\cm^2}$ \autocite{Hobold2023}. This agreement validates Eq.~\ref{eq:j0p} and provides a direct experimental method to determine $\delta$ using the measured apparent rate ($j_0^p$), intrinsic rate ($j_0^{\text{UME}}$), and electrolyte properties.

In classical electrochemical theory, the applied voltage diverges as the applied current density $j_{\text{app}}\rightarrow j_{\text{lim}}$ \autocite{Bazant2005}, implying that $j_{\text{lim}}$ sets an absolute upper bound on the operating current density, i.e., $j_{\text{app}}<j_{\text{lim}}$. However, experiments on the studied system frequently report $j_{\text{app}}$ exceeding this limit. This discrepancy can be reconciled by accounting for SEI breakdown \autocite{Boyle2022}. As depicted in Fig.~\ref{fig:schematic}C, localized high current densities and associated mechanical stress promote SEI thinning, microcracking, and electrolyte infiltration, leading to an effective increase in SEI diffusivity and concentration. These processes collectively reduce the effective SEI parameter $\delta$, thereby enhancing $j_{\text{lim}}$. Experimentally, a critical current density of $4.5~\si{\mA/\cm^2}$ marks the onset of significant SEI breakdown \autocite{Boyle2022}. Consistently, our model shows that the limiting current causes $\delta$ to drop sharply from 386 to below 77 at this critical value, enabling observable SEI degradation. To quantify the breakdown, we introduce the dimensionless SEI breakdown parameter $\beta$, defined as
\[\beta = j_{\text{lim}}^c \frac{\mathrm{d} \delta(j)}{\mathrm{d} j},\]
which captures the sensitivity of $\delta$ to local current fluctuations. Given the inherent stochasticity and heterogeneity of SEI, we do not assume a universal breakdown curve. Instead, $\beta$ is treated as a local, environment-dependent quantity that characterizes the likelihood and severity of breakdown at a given $j$-$\delta_0$ state. The macroscopic breakdown process is viewed as a superposition of such local events, dominated by the most unstable case (most negative $\beta$). The system's state, which depends on the local environment, can lie anywhere within the gray region (Fig.~\ref{fig:schematic}D) bounded by $j_{\text{lim}}$. Note that $\beta$ can also describe SEI thickening.

In this work, we employ linear stability analysis \autocite{Sekerka2015,Zhang2024,Zhang2025} to analyze morphological instability of the lithium electrode. As illustrated in Fig.~\ref{fig:schematic}A, the interface is assumed to have a small perturbation of the form $h(x,t)=\varepsilon e^{ik x + \Sigma t}$, where $i= \sqrt{-1}$, $\varepsilon\ll 1$, $k$ is the wavenumber, and $\Sigma$ denotes the growth rate. Negative $\Sigma$ indicates stability where the perturbation shrinks, while positive $\Sigma$ signifies instability where the perturbation grows. To assess relative growth compared to a planar interface, we normalize the growth rate by $j_{\text{app}}$ to define a scaled dimensionless growth rate $\sigma = F \Sigma/(V_m j_{\text{app}})$. To facilitate analysis, we introduce dimensionless variables denoted by a tilde: the wavenumber $\tilde{k}=kL$, current density $\tilde{j} = j/j_{\text{lim}}^c$, and overpotential $\tilde{\eta} = F \eta/RT$.

The general expression for the scaled growth rate is derived (see Supporting Information) for general charge transfer kinetics $\tilde{j}(\tilde{\eta})$ and solvation/desolvation dynamics $\tilde{j}(\tilde{\eta}_{\text{solv}})$:
\begin{equation}
  \label{eq:sigma:general}
  \sigma(\tilde{k}) = \frac{1- \tilde{k}^2 \text{Ca} \frac{1-\tilde{j}_{\text{app}}\delta_m}{4\tilde{j}_{\text{app}}}}{\frac{1-\tilde{j}_{\text{app}}\delta_m}{-4\tilde{j}_{,\tilde{\eta}}} + \frac{\delta+\tilde{j}_{\text{app}}\beta}{2}\exp\left(-\frac{\tilde{\eta}_{\text{solv}}}{2}\right) - \frac{1-\tilde{j}_{\text{app}}}{4}\tilde{\eta}_{\text{solv},\tilde{j}} + B(\tilde{k})},
\end{equation}
where $\tilde{j}_{,\tilde{\eta}}=\mathrm{d}\tilde{j}/\mathrm{d}\tilde{\eta}$, $\tilde{\eta}_{\text{solv},\tilde{j}}=\mathrm{d}\tilde{\eta}_{\text{solv}}/\mathrm{d}\tilde{j}$, $\delta_m = 1+\delta \exp\left(-\tilde{\eta}_{\text{solv}}/2\right)$, $B(\tilde{k})$ defined in Eq.~S68, $\text{Ca}=V_m \gamma/RTL$ is the capillary number, and $\gamma$ the total interfacial energy of the electrode-SEI and SEI-electrolyte interfaces. In the case of symmetric Butler-Volmer charge transfer kinetics and linear solvation/desolvation dynamics, the scaled growth rate simplifies to:
\begin{equation}
  \label{eq:sigma}
  \sigma(\tilde{k}) = \frac{1- \tilde{k}^2 \text{Ca} \frac{1-\tilde{j}_{\text{app}}\delta_m}{4\tilde{j}_{\text{app}}}}{\frac{1-\tilde{j}_{\text{app}}\delta_m}{2\sqrt{\tilde{j}_{\text{app}}^2 + 4 \tilde{j}_0^2}} +  \frac{\delta+\tilde{j}_{\text{app}}\beta}{2}\exp\left(\frac{\tilde{j}_{\text{app}}}{2\tilde{j}_{0,\text{solv}}}\right) + \frac{1-\tilde{j}_{\text{app}}}{4 \tilde{j}_{0,\text{solv}}} + B(\tilde{k})}.
\end{equation}

This expression reveals that the stability of the lithium electrode interface is governed by six dimensionless parameters: the applied current density $\tilde{j}_{\text{app}}=j_{\text{app}}/j_{\text{lim}}^{\text{c}}$, the capillary number $\text{Ca}$ \autocite{Khoo2019}, the charge transfer rate $\tilde{j}_0 = j_0/j_{\text{lim}}^c$ (classical Damk\"ohler number), the solvation rate $\tilde{j}_{0,\text{solv}} = j_{0,\text{solv}}/j_{\text{lim}}^c$, the SEI parameter $\delta$, and the SEI breakdown parameter $\beta$. These parameters collectively determine the onset and evolution of morphological instabilities.

\begin{figure*}[!t]
  \centering
  \includegraphics[width=1.0\linewidth]{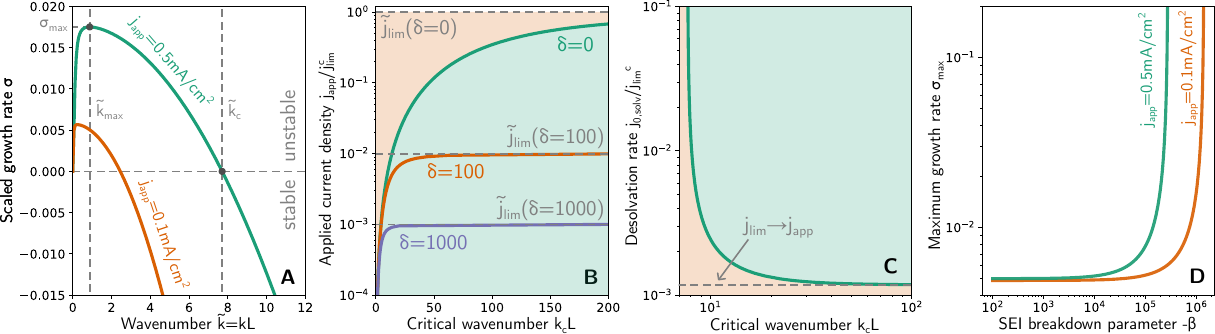}%
  \caption{Stability results. (A) Dispersion curves for two applied current densities. The critical wavenumber $\tilde{k}_c$ defines the stability boundary ($\sigma=0$) and the maximum growth rate $\sigma_{\text{max}}$ occurs at $k_{\text{max}}$. (B) Impact of the dimensionless applied current density $j_{\text{app}}/j_{\text{lim}}^c$ on the dimensionless critical wavenumber $k_cL$ for three values of the SEI parameter $\delta$. (C) Effect of the dimensionless desolvation rate ($j_{0,\text{solv}}/j_{\text{lim}}^c$) on the stability boundary $\tilde{k}_c$. Dashed lines in (B) and (C) indicate the approaching proximity between the applied current and the limiting current. (D) Effect of the SEI breakdown parameter ($\beta$) on the maximum growth rate for two applied current densities.}\label{fig:stability}
\end{figure*}

\begin{figure}[!t]
  \centering
  \includegraphics[width=1.0\linewidth]{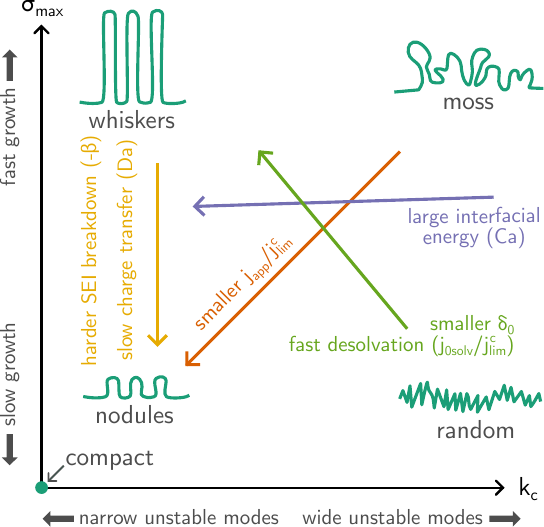}%
  \caption{Qualitative morphology instability map for lithium electrodeposition. This map is defined by the critical wavenumber $k_c$ and the maximum growth rate $\sigma_{\text{max}}$. Different regions correspond to distinct morphologies: Mossy dendrites (large $k_c$, large $\sigma_{\text{max}}$), whisker dendrites (small $k_c$, large $\sigma_{\text{max}}$), nodule-like dendrites (small $k_c$, small $\sigma_{\text{max}}$), and random, small-amplitude morphologies (large $k_c$, small $\sigma_{\text{max}}$). The region characterized by minimal $k_c$ and $\sigma_{\text{max}}$ is associated with a compact morphology. Arrows illustrate the trajectory and influence of the six dimensionless governing parameters on the morphology.}\label{fig:morphology}
\end{figure}

In Fig.~\ref{fig:stability}A, the dispersion curves illustrate the scaled growth rate of interface perturbations as a function of $\tilde{k}$. The curve crosses zero at the critical wavenumber $\tilde{k}_c$ that defines the stability boundary: $\sigma(\tilde{k})=0$. The dispersion curve reaches a maximum growth rate $\sigma_{\text{max}}$ at $\tilde{k}_{\text{max}}$, which indicates the most unstable mode and thus the dominant mechanism driving morphological instability. Notably, $\sigma=0$ at $\tilde{k}=0$, consistent with a planar interface, where no relative growth occurs compared to the unperturbed state. Together, $k_c$ and $\sigma_{\text{max}}$ characterize the shape of the dispersion curve and provide a predictive framework for the electrodeposition morphology (Fig.~\ref{fig:morphology}). $k_c$ governs the characteristic spacing and curvature of the dendritic pattern, and $\sigma_{\text{max}}$ correlates with the dendrite tip velocity. A large $k_c$ implies a broad spectrum of unstable modes, promoting complex, irregular morphologies in the subsequent nonlinear growth. Low growth rates yield a random, small-amplitude structure that favors porous morphologies upon cycling. With an increasing growth rate, the morphology evolves into mossy dendrites. Conversely, a smaller $k_c$ restricts instability to a narrow mode range, resulting in simpler, more periodic morphologies, such as nodule-like structures at low $\sigma_{\text{max}}$ and whiskers at high $\sigma_{\text{max}}$. Increasing $j_{\text{app}}$ shifts $\tilde{k}_c$ and $\sigma_{\text{max}}$ to higher values, thereby promoting more complex and rapidly developing patterns.

The critical wavenumber $\tilde{k}_c$ can be derived as
\begin{equation}\label{eq:kc:dendrite}
  \tilde{k}_c = \sqrt{\frac{4}{\text{Ca}}\frac{\tilde{j}_{\text{app}}}{1 - \tilde{j}_{\text{app}} \delta_m}}.
\end{equation}
Comparing Eq.~\ref{eq:cSEI0} with the denominator of Eq.~\ref{eq:kc:dendrite} reveals that $\tilde{k}_c$ diverges as $j_{\text{app}}$ approaches $j_{\text{lim}}$, as illustrated by the horizontal dashed asymptotes in Figs.~\ref{fig:stability}B-C. In this case, even the short-wavelength perturbations are unstable. This explains why higher salt concentration (higher $j_{\text{lim}}$) promotes nodule-like morphologies \autocite{PlazaRivera2025}. The dependence of $k_c$ on other parameters is detailed in the Supporting Information. Notably, $\tilde{j}_0$ and $\beta$ do not affect $\tilde{k}_c$.

The influence of all six parameters on $\sigma_{\text{max}}$ is systematically analyzed (Fig.~\ref{fig:stability}D; with full parametric results in Fig.~S2). In contrast to the divergent behavior observed in Fig.~\ref{fig:stability}D, $\sigma_{\text{max}}$ remains finite for all other parameters (Fig.~S2), indicating a strong influence of $\beta$ on $\sigma_{\text{max}}$. This confirms the crucial impact of SEI integrity on stability, aligning with experimental observations \autocite{Mi2025}. The $\beta$ value may depend on the SEI type: layered vs. mosaic-like \autocite{Hobold2024}. For $\beta=0$, Fig.~S3 shows positive correlation between $\sigma_{\text{max}}$ and the measured CE \autocite{Hobold2023}. This discrepancy is likely due to the unknown $\beta$ value. The dominant effect of $\beta$ on $\sigma_{\text{max}}$ may therefore overwhelm the influence of other parameters, emphasizing the need for future experimental quantification. For fixed $\beta$, we observe a positive correlation between $\sigma_{\text{max}}$ and $j_0^p$ (Fig.~S4). This is consistent with the principle that slower reaction promotes a reaction-limited regime, thereby suppressing diffusion-related instabilities. While this mechanism suggests that a larger $\delta$ and a slower desolvation (smaller $j_{0,\text{solv}}$) is preferred (based on Fig.~\ref{fig:jlim}B), these parameters simultaneously affect $j_0^p$ and $j_{\text{lim}}$. To characterize the combined effects, we introduce an apparent Damk\"ohler number 
\begin{equation}
  \label{eq:Da}
  \text{Da}^p=\frac{j_0^p}{j_{\text{lim}}},
\end{equation}
analogous to the classical Damk\"ohler number $\text{Da}=j_0/j_{\text{lim}}^c$ in SEI-free systems \autocite{Khoo2019,Cogswell2015}. $\text{Da}^p$ quantifies the relative importance of reaction kinetics (governed by $j_0^p$) and mass transport (governed by $j_{\text{lim}}$). High $\text{Da}^p$ implies a fast surface reaction relative to diffusion (the system is diffusion-limited), leading to rapid ion depletion, steep concentration gradients, and a destabilized interface that favors dendrite growth. Lower $\text{Da}^p$ shifts the system toward reaction control, favoring stable, planar deposition. Fig.~S3 shows $\text{Da}^p$ correlates well with the measured CE. Compared to previously proposed metrics, such as $j_0^p/FcD \propto j_0^p/j_{\text{lim}}^c$ \autocite{Hobold2021} and $\text{cT} \propto j_{\text{lim}}^{\text{SEI}}$ \autocite{Liu2025}, $\text{Da}^p$ offers a more complete picture of the interplay between reaction and diffusion while incorporating the effects of SEI and desolvation. Notably, higher $j_0^p$ does not contradict a lower $\text{Da}^p$, as seen in Figs.~\ref{fig:jlim}B and C. Recent experiments \autocite{Kim2025} reveal improved performance in DME:TTE electrolytes with reduced solvation strength. Weaker solvation correlates with faster desolvation kinetics and lower ionic diffusivity, lowering $\delta$ and increasing $j_{0,\text{solv}}/j_{\text{lim}}^c$. This shifts the system toward the upper-left region of Fig.~\ref{fig:jlim}C, where Da$^p$ is minimized, consistent with the observed improvement in electrochemical performance. Optimal battery design involves a stability-performance trade-off: minimizing $\sigma_{\text{max}}$ requires large $\delta$ and small $j_{0,\text{solv}}$, while achieving high operating current (increased $j_{\text{lim}}$) and reducing $k_c$ demands small $\delta$ and large $j_{0,\text{solv}}$.

Fig.~\ref{fig:morphology} systematically maps the influence of six dimensionless parameters on electrodeposition morphology. Increasing $j_{\text{app}}$ simultaneously raises $k_c$ and $\sigma_{\text{max}}$, thereby promoting more complex morphologies. While the SEI breakdown parameter $\beta$ and the charge transfer rate $j_0$ do not alter $k_c$, they strongly affect $\sigma_{\text{max}}$, indicating that accelerated SEI breakdown and faster charge transfer amplify instability. The capillary number Ca has a minor influence on $\sigma_{\text{max}}$, but an increased interfacial energy strongly suppresses $k_c$, narrowing the range of unstable wavelengths. The interplay between the SEI parameter $\delta$ and the desolvation rate $\tilde{j}_{0,\text{solv}}$ reveal a fundamental trade-off between $\sigma_{\text{max}}$ and $k_c$, which can be quantified by $\text{Da}^p$. This work provides a comprehensive understanding of the coupled mechanisms governing interfacial stability, guiding the development of stable, high-performance electrodeposition systems essential for next-generation energy storage. Chemo-mechanical interactions are critical for interfacial stability. In the current model, stress-induced SEI failure is captured by the SEI breakdown parameter $\beta$. While we adopt a homogeneous SEI model with averaged properties for simplicity, the multiscale framework is inherently extensible to heterogeneous SEI models that capture the complexity of SEI nanostructure, evolution and the effects of local stress-induced failure. Future work will extend this framework to nonlinear interfacial evolution using phase-field modeling \autocite{Zhang2023}, enabling the incorporation of detailed SEI models and full chemo-mechanical coupling.

\section*{Acknowledgements}
The authors thank support from Office of Naval Research under grant N00014-20-1-2327 and financial assistance Award 70NANB14H012 from the U.S. Department of Commerce, National Institute of Standards and Technology as part of the Center for Hierarchical Materials Design (CHiMaD). We thank Jeffrey Lopez for his valuable comments.

\section*{Supporting Information}
The Supporting Information is available at \url{https://pubs.acs.org/doi/10.1021/acsenergylett.5c03690}.

\printbibliography[title={\sffamily\bfseries References}]

\end{document}